\documentclass[%
 reprint,
 amsmath,amssymb,
prl,
]{revtex4-2}

\usepackage{graphicx}
\usepackage{dcolumn}
\usepackage{bm}
\usepackage{hyperref}

\begin{document}

\title{Dynamics of the Transverse Optical Flux in Random Media}

\author{Yuchen Ke$^{1}$}
\author{Nandini Bhattacharya$^{1}$}
\author{Fabian Maucher$^{1}$}
 \email{f.maucher@tudelft.nl}
\affiliation{
$^1$Department of Precision and Microsystems Engineering, Faculty of Mechanical Engineering, Delft University of Technology, 2628 CD Delft, The Netherlands
}

\begin{abstract}
We study the evolution of the kinetic energy (or gradient norm) of an incident linearly polarized monochromatic wave propagating in correlated random media. We explore the optical flux transverse to the mean Poynting flux at the paraxial-nonparaxial (vectorial) transition along with vortex counting. Here, by paraxial–nonparaxial transition we mean a gradual loss of validity of the paraxial approximation such that it is necessary to solve Maxwell-consistently employing the dyadic Green's function. 
The vortex number appears to increase approximately with a cubic root of the propagation distance for sufficiently small correlation length. Furthermore, a kink appears in nucleation rate at the position of maximum scintillation upon increasing correlation length. A driven steady state is reached due to the filtering of evanescent waves upon propagation.  Finally, we present the spectrum of the incompressible kinetic energy and how it evolves from the paraxial case to that of a (nonparaxial) random field. 
\end{abstract}

\maketitle

\section{I. Introduction}
The propagation of light in a medium causes microscopic superposition between the incident field and the atomic or molecular dipole fields~\cite{Huffman:book}. Optical media are usually dense; i.e., the interparticle separation is usually much smaller than the wavelength; thus, each particle sees the secondary fields of a large neighborhood of particles. This gives rise to an overall macroscopic response in the form of a refractive index of the medium. 
Heterogeneity of the media leads to modulations of the refractive index~\cite{Vynck:RMP:2023}, which to lowest order can be described as correlated random perturbations. 
The propagation of light in random media is complex, as it often involves taking into account multiple scattering and material properties~\cite{Huffman:book,Ishimaru:book,Nieuwenhuizen:RMP:1999,Carminati:book:2021}. 
The effects on the propagation of light in random media became further accessible due to the availability of coherent monochromatic sources. 
Intriguing phenomena emerge such as atmospheric scintillation~\cite{Andrews:JOSAA:1999}, branching of light~\cite{Patsyk:nature:2020,Chang:NatComm:2024}, or Anderson localization~\cite{Anderson:PhysRev:1958,Wiersma:Nature:1997,Segev:Nature:2007, Lagendijk:PhysToday:2009,Cao:NatPhys:2023}. Through multichannel interference~\cite{Dorokhov:SSC:1984}, light propagation in random medium can act focusing like a lens~\cite{vellekoop:ol:2007,Bertolotti:nature:2012}. Furthermore, spectrometers can be based on random media ~\cite{Cao:JoO:2017,Cao:NatPhot:2013}. 
Upon traveling through random media, vortices~\cite{Berry:ProcA:1974} nucleate in light. %. 
Optical vortices are quantized, pointlike null regions of intensity surrounded by azimuthal phase ramp, and thus determine the azimuthal flow of light. Vortices can be thought of as a singular skeleton~\cite{Soskin:ProgOpt:2001,Dennis:ProgOpt:2009} for the speckle pattern, carrying information that is complementary to the amplitude, employed for, e.g., multiplexing~\cite{Wang:NatPhot:2012}. Vortices permit advanced trapping control~\cite{Grier:Nat:2003}, and the associated momentum can even be entangled on the single-photon level~\cite{Zeilinger:Nature:2001}.
Quantized vortices also occur in ultracold quantum fluids~\cite{Barenghi:book:2001}, where instead of propagating through a random medium they can be nucleated via optical stirring~\cite{Cornell:PRL:1999} or magnetic quenching leading to quantum turbulence~\cite{Bagnato:PRL:2009}. Vortex statistics of optical beams after a planar speckle field have been extensively studied both theoretically~\cite{Berry:JphysA:1978,Freund:OptComm:1993,Freund:JoptSoc:94} and experimentally~\cite{Shkukov:JETP:1981,Baranova:83}. 
Furthermore, local properties such as anisotropy of the vortex core structure and circular flux current have been successfully predicted~\cite{Dennis:Proc:2000} and experimentally verified~\cite{Zhang:PRL:2007, Zhang_Genack:PRL:2007}. 
The interplay between optical speckle and vortices has been explored in volumetric random fields in homogeneous media~\cite{Dennis:Proc:2000, Padgett:PRL:2008}, and the ensemble dynamics of point vortices in speckle fields has been studied~\cite{Staliunas:OptComm:1995}. 

\begin{figure}[htbp]
\includegraphics[width=0.8\columnwidth]{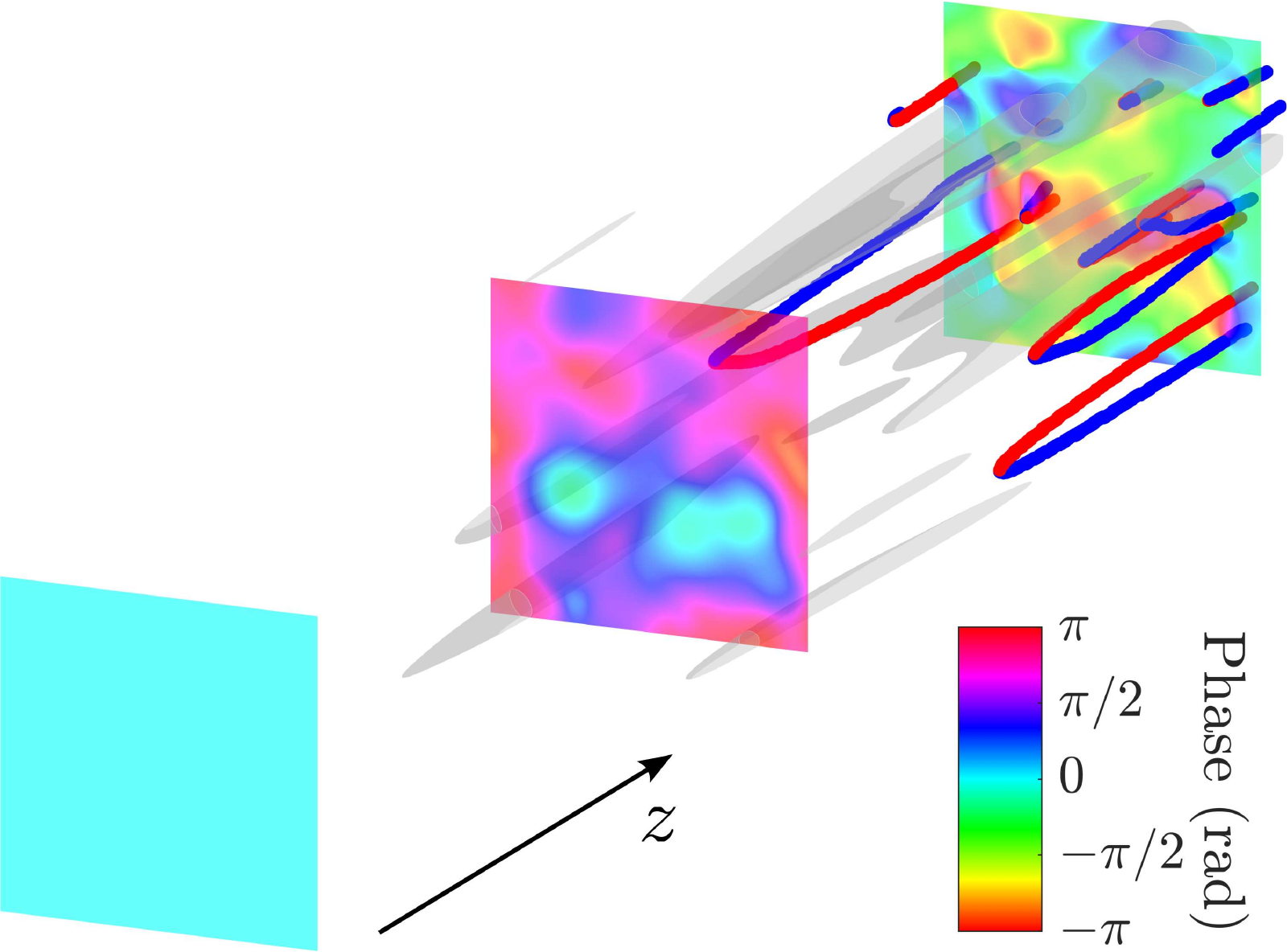}
\caption{\label{fig:fig1} Propagation of an initially linearly polarized plane wave in a random medium. The planes depict the phase, the lines correspond to vortex lines [red (blue) topological charge $1$ ($-1$)], and the gray iso-surface depicts the intensity, showing speckle formation. }
\end{figure}

In this work, we aim at further understanding and statistically characterizing 
light as it evolves upon the propagation of monochromatic waves in random media, as visualized in Fig.~\ref{fig:fig1}. We consider randomness to be fully determined by amplitude and correlation length of the refractive index~\cite{Vynck:RMP:2023}. 

The randomness of the medium is gradually imprinted onto the electric field while it propagates through the medium and evolves from a plane wave to a fully developed speckle field. Furthermore, the inhomogeneity of the medium couples the polarization components, converting the initial scalar field into a vectorial one. 
Thus, the evolution of the wave in a medium cannot be characterized solely by 
statistical approaches that globally describe the light as Gaussian random waves based on the statistics of the medium. 
In other words, we consider the transition from quasilinearly polarized, paraxial domains to regions of nonparaxial and vectorial light. 

We approach this problem by studying the ``kinetic energy" transverse to the propagation direction, i.e., the gradient norm of the electric field, as well as vortex counting upon propagation. For the early propagation, we compare the numerics with a paraxial analytical expression for the kinetic energy. We then characterize the dynamics by the number of vortices per unit area as a function of propagation distance in the medium. Finally, we consider the energy spectrum - i.e., incompressible kinetic energy as a function of the wave vector -  borrowing methods from quantum turbulence theory~\cite{Bradley:PRX:2012} as considered for optics recently in Ref.~\cite{Glorieux:PRA:2023}. 

\section{II. Modeling of Light in Random Media}

From Maxwell's equations, it is simple to find that the wave propagation of the electrical field ${\bm  E}({\bm  r})$ in a linear, non-magnetizable medium is governed by the following vector wave equation:
\begin{equation}
    \nabla \times \nabla\times{\bm  E}({\bm  r}) - k({\bm  r})^2 {\bm  E}({\bm  r})     = {\bm \xi}({\bm r}). 
\label{eq:vector_Helmholtz}
\end{equation}
Here, ${\bm  r}=(x,y,z)$ and ${\bm k}({\bm  r})$ is the wave vector with modulus $k({\bm  r})= 2\pi n({\bm  r})/\lambda$, where $\lambda$ is the wavelength in vacuum. The refractive index is given by $n({\bm  r})=n_0+\eta({\bm  r})$~\cite{Fante:ProcIEEE:1975}, where $n_0$ is the constant background refractive index and $\eta({\bm  r})$ the spatially dependent deviation. 
The vector ${\bm \xi}({\bm r})$ corresponds to the source of the field, which is assumed to be zero everywhere apart from the input plane at $z=0$. 
We assume that the direction of the mean Poynting flux is equivalent to the $z$ direction. 
Furthermore, we presuppose $\epsilon$ to be a $\delta$-correlated Langevin Gaussian
distributed random noise as a function of the Cartesian coordinate $\bm r$. 
We introduce the ensemble average over different stochastic realizations $f_i$ as $\langle f\rangle = \lim_{N\rightarrow\infty}\frac{1}{N}\sum_{i=1}^N f_i$.
Then, the noise satisfies $\langle\epsilon\rangle=0$ and its correlation is given by $\langle\epsilon({\bm  r})\epsilon({\bm  r}^\prime)\rangle = \kappa^2\delta({\bm r}-{\bm r}^\prime)$, where $\kappa$ is the coupling strength characterizing the amplitude of the noise. By convolving the white noise $\epsilon$ with a normalized Gaussian function $g({\bm  r}) = \mathrm{e}^{-\frac{{\bm r}^2}{\sigma^2}}/(\pi^{3/2}\sigma^3)$,
we obtain a Gaussian correlated noise $\eta({\bm  r})=\int g({\bm r-\bm r}^\prime) \epsilon({\bm r}^\prime) d{\bm r}^\prime$, i.e., 
\begin{equation}
    \langle \eta({\bm r}) \eta({\bm r}^\prime) \rangle = C({\bm r}-{\bm r}^\prime),
\end{equation}
where we introduced the correlation function $C$,
\begin{equation}
    C({\bm r}-{\bm r}^\prime) = \frac{\kappa^2}{{(2\pi\sigma^2)}^{3/2}} \mathrm{e}^{-({\bm r}-{\bm r}^\prime)^2/(2 \sigma^2)}.
\end{equation}

Hence, Eq.~(\ref{eq:vector_Helmholtz}) has two free parameters $\sigma/\lambda$ and $\kappa/\lambda^{3/2}$.
To propagate the wave in the random medium, we use the modified Born series approach described in Refs.~\cite{osnabrugge:JCompPhys:2016, osnabrugge:OpEx:2021}, which unlike the ``unmodified" Born series extends convergence substantially. As breakup condition, we impose that the relative difference of the norm of two consecutive solutions has to be smaller than $\varepsilon=10^{-7}$. To check whether we found an actual solution to Eq.~(\ref{eq:vector_Helmholtz}), we take the norm of the equation and demand that this is smaller than $\delta=10^{-5}$. We assume the source to be a linearly polarized plane wave with polarization along the $x$ axis. We typically use 64 cores for simulating a grid of $N_x=N_y=256-512$ and $N_z=512-8192$ points and roughly $100-40000$ iterations. Without loss of generality, the background refractive index of all media is set to $n_0 = 1.5$.
We use periodic boundary conditions in the ${\bm r}_\perp=(x,y)$ plane transverse to the mean Poynting flux to mimic the thermodynamic limit - i.e., an infinite transverse plane $\mathcal{A}$ - and absorbing boundary layer in the $z$ direction parallel to the mean Poynting flux, implemented as an antireflection layer with acyclic convolution~\cite{osnabrugge:OpEx:2021}. 

We use this framework to investigate the propagation. For convenience, we restrict our consideration to the original  $x$ polarization and define the transverse kinetic energy per unit area as 
\begin{equation}
    E_{\text{kin}}(z)=\lim_{\mathcal{A}\rightarrow\infty}\frac{1}{\mathcal{A}}\int_\mathcal{A} |\nabla_\perp E_x({\bm r}_\perp,z) |^2 d^2r_\perp.
\end{equation}

\section{III. Characterization of the Propagation Dynamics}

In the following subsections, we would like to characterize the different regimes of light propagation in the random medium. For that matter, we first describe the initial dynamics, where propagation is only weakly perturbed, employing the gradient norm. After that, we characterize the full dynamics by vortex counting and finally by studying the kinetic energy spectrum. 

\subsection{A. Initial paraxial dynamics}

In this subsection, we focus on the initial dynamics only and aim at finding analytical expressions to characterize it. 
Let us start with deriving the basic equations describing this weakly modulated regime. 
Gauss's law of Maxwell's equations without sources is given by $\nabla\cdot (\epsilon {\bm E})=0$. We see that if the permittivity $\epsilon$ varies spatially slowly on the scale of the wavelength, 
i.e., $|\nabla\epsilon/\epsilon|\ll k_0$, 
we can approximate $\nabla\cdot{\bm E}\approx 0$. This reduces the wave equation [Eq. (1)] to the vector  Helmholtz equation. In case of our incident $x$-polarized wave, its propagation for $z>0$ is given by 
\begin{equation}
    \nabla^2E_x({\bm  r}) + k({\bm  r})^2E_x({\bm  r}) = 0.
\end{equation}
Assuming that the slowly varying envelope $U$ of $E_x=U\mathrm{e}^{\mathrm{i}k_0z}$ with $k_0 = \frac{2\pi}{\lambda}n_0$ satisfies $|\partial_{zz}U|\ll |k_0\partial_zU|$, we can perform the slowly varying envelope approximation. 
Linearization in $\eta$ yields the paraxial wave equation for $U$: 
\begin{equation}
     2\mathrm{i}k_0\partial_z U = -\nabla^2_\perp U - \frac{4\pi}{\lambda}\eta k_0U.
\label{eq:paraxial_wave_eq}
\end{equation}
We define the kinetic energy $E_\text{kin}$ at each $z$ plane as 
\begin{equation}
    E_\text{kin}(z)= \int|\nabla_\perp U({\bm r}_\perp,z) |^2 d^2r_\perp,  
    \label{eq:define_Ekin}
\end{equation}
such that $E_\text{kin}$ satisfies $\frac{\delta E_\text{kin}}{\delta U^*} = 2\mathrm{i}k_0\partial_z U$ when $\eta \equiv 0$. Here, $U^*$ denotes the complex conjugate of $U$. The spatial derivative of the averaged kinetic energy along the direction of wave propagation is
\begin{equation}
    \partial_z \langle E_\text{kin}  \rangle= \mathrm{i}\frac{2\pi}{\lambda}\left \langle \int \eta(U^*\nabla^2_\perp U - U\nabla^2_\perp U^*) d^2r_\perp \right \rangle.
\label{eq:Ekin_z_derivative}
\end{equation} 
Since the rapid oscillating term $\mathrm{e}^{\mathrm{i}k_0z}$ is factored out, $U$ varies on a much longer length scale in the $z$ direction than in the $x$ and $y$ directions. 
As a result, the ratio between correlation length $\sigma$ of the fluctuation of refractive index and the variation length scale in $z$ direction is much smaller than the ratio in the $(x,y)$ plane. 
Therefore, it appears reasonable to apply Markov approximation; i.e., assume that the refractive index fluctuation is delta correlated in $z$ direction. With that, the correlator of $\eta$ becomes
\begin{equation}
    \langle \eta({\bm r}) \eta({\bm r}^\prime) \rangle = C_\perp({\bm r}_\perp-{\bm r}_\perp^\prime)\delta(z-z'), 
\label{eq:autocorrelation_2}
\end{equation}
and the corresponding transverse correlation function is 
\begin{equation}
    C_\perp({\bm r}_\perp-{\bm r}_\perp^\prime) = \frac{\kappa^2}{2\pi\sigma^2}\mathrm{e}^{-({\bm r}_\perp-{\bm r}_\perp^\prime)^2/(2 \sigma^2)}.
\end{equation}

By applying the Furutsu-Donsker-Novikov formula~\cite{novikov1965functionals, furutsu1963statistical}
\begin{equation}
    \left\langle \eta U \right\rangle = \int_{-\infty}^z\int \left\langle \frac{\delta U({\bm r}_\perp, z)}{\delta\eta({\bm r}_\perp^\prime, z^\prime)}\right\rangle\left\langle\eta({\bm r}_\perp, z)\eta({\bm r}_\perp^\prime, z^\prime)\right\rangle d^2r_\perp^\prime dz^\prime
\end{equation}
to Eq.~(\ref{eq:Ekin_z_derivative}), and following similar derivation steps in Ref.~\cite{maucher2012stability}, we obtain
\begin{align}
    \partial_z \langle E_\text{kin}  \rangle &= -\frac{4\pi^2}{\lambda^2}\nabla^2_\perp C_\perp({\bm r}_\perp)\bigg|_{{\bm r}_\perp=0}\left\langle\int|U|^2d^2r_\perp\right\rangle\nonumber \\
    &= -\frac{4\sqrt{2}\pi^{5/2}\sigma}{\lambda^2}\nabla^2_\perp C({\bm r}_\perp, 0)\bigg|_{{\bm r}_\perp=0}\left\langle\int|E_x|^2d^2r_\perp\right\rangle\nonumber \\
    &= \frac{4\pi\kappa^2}{\sigma^4\lambda^2}\mathcal{I}_x.
    \label{eq:slope_EK}
\end{align}

The evolution of the ensemble-averaged kinetic energy is a linear function of the propagation distance $z$. If we do not ensemble average, there is a Brownian motion around that linear trend.
Here, we introduced the intensity $\mathcal{I}_i$ or average power per area $\mathcal{A}$ as $\mathcal{I}_i = \lim_{\mathcal{A}\rightarrow\infty}\frac{1}{\mathcal{A}}\int_\mathcal{A} |E_i|^2 d^2r_\perp$. 

\begin{figure}[hbt]
\includegraphics[width=0.48\textwidth]{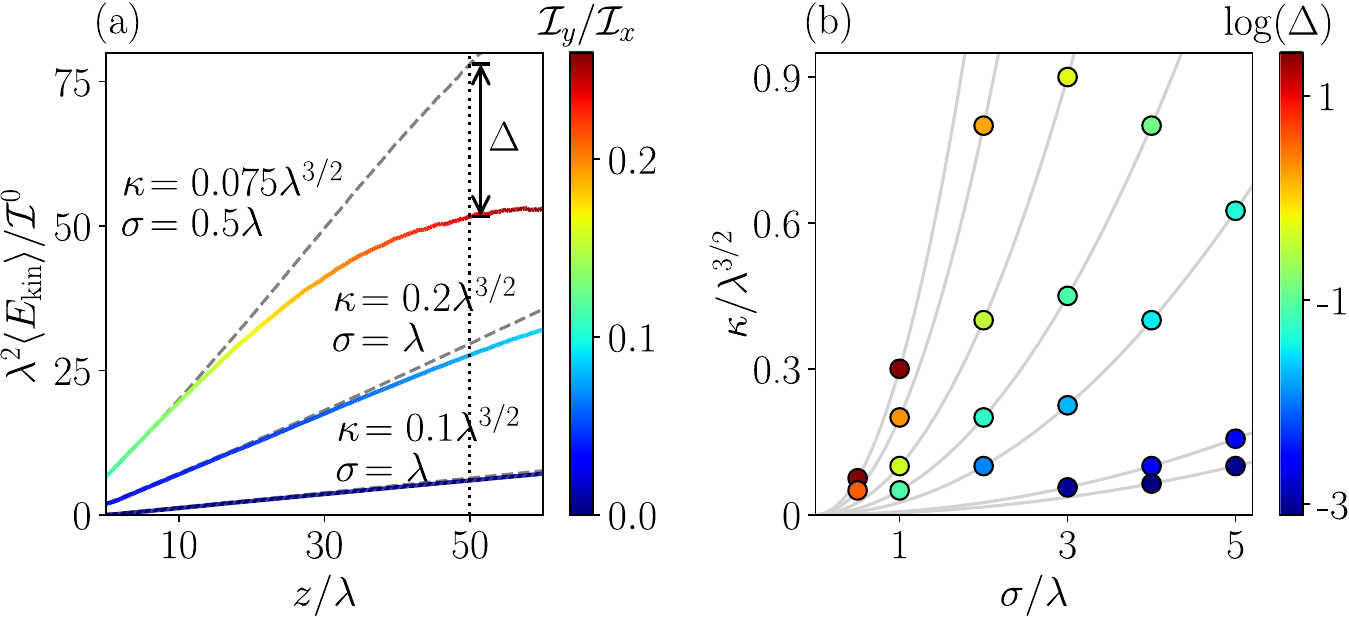}
\caption{\label{fig:Hamiltonian} (a) Comparison of numerical simulations (solid lines) and analytical predictions (dashed lines) for the evolution of the ensemble-averaged kinetic energy at the initial stage. $\Delta$ denotes the difference evaluated at $z = 50\lambda$. The color of the solid lines is given by $\mathcal{I}_y/\mathcal{I}_x$ and, therefore, indicates how vectorial the light becomes upon propagation. (b) The values of $\Delta$ for various $\sigma$ and $\kappa$, represented on a logarithmic scale.
The gray lines represent isolines where $\kappa^2\lambda/\sigma^4$ is constant. The ensemble average was performed over ten realizations.}
\end{figure}

The comparison between the analytical prediction and numerics is shown in Fig.~\ref{fig:Hamiltonian}, where the quantity $\mathcal{I}^0=\mathcal{I}(\eta=0)$ was introduced for normalization purposes. The simulations are performed in a domain of size
$L_x=L_y=80\lambda$ and $L_z=120\lambda$, discretized on a uniform grid with $N_x=N_y=512$ and $N_z=768$. 
From the paraxial consideration [Eq.~(\ref{eq:slope_EK})], one would expect a scaling $\sim \kappa^2 \mathcal{I}_x/\sigma^4\lambda^2$ of the slope. Yet, evidently, this scaling is violated quickly once the propagation becomes nonparaxial and vectorial, leading to a significant mismatch. 
To illustrate that, we colored the numerical data with $\mathcal{I}_y/\mathcal{I}_x$.
The curves that start with finite initial values are due to backscattering. 
In other words, the field at the plane of the source is not necessarily proportional to the source. 
The difference $\Delta$ evaluated as a function of $\kappa$ and $\sigma$ is illustrated by the colormap in Fig.~\ref{fig:Hamiltonian}(b) in logarithmic scale. 
We see that decreasing $\sigma$ or increasing $\kappa$ results in the increase of the difference $\Delta$. This can be expected, as both lead to a deviation from paraxiality. 
To further qualitatively interpret the result, consider the gray lines representing isolines, where the slope is predicted to be equal. 
We observe that the difference $\Delta$ is roughly equal on these lines. 
Subsequently, the increase in kinetic energy leads to the nucleation of vortex pairs, which will be explored in the next subsection. 

\subsection{B. Vortex nucleation rate}

\begin{figure}[htbp]
\includegraphics[width=0.48\textwidth]{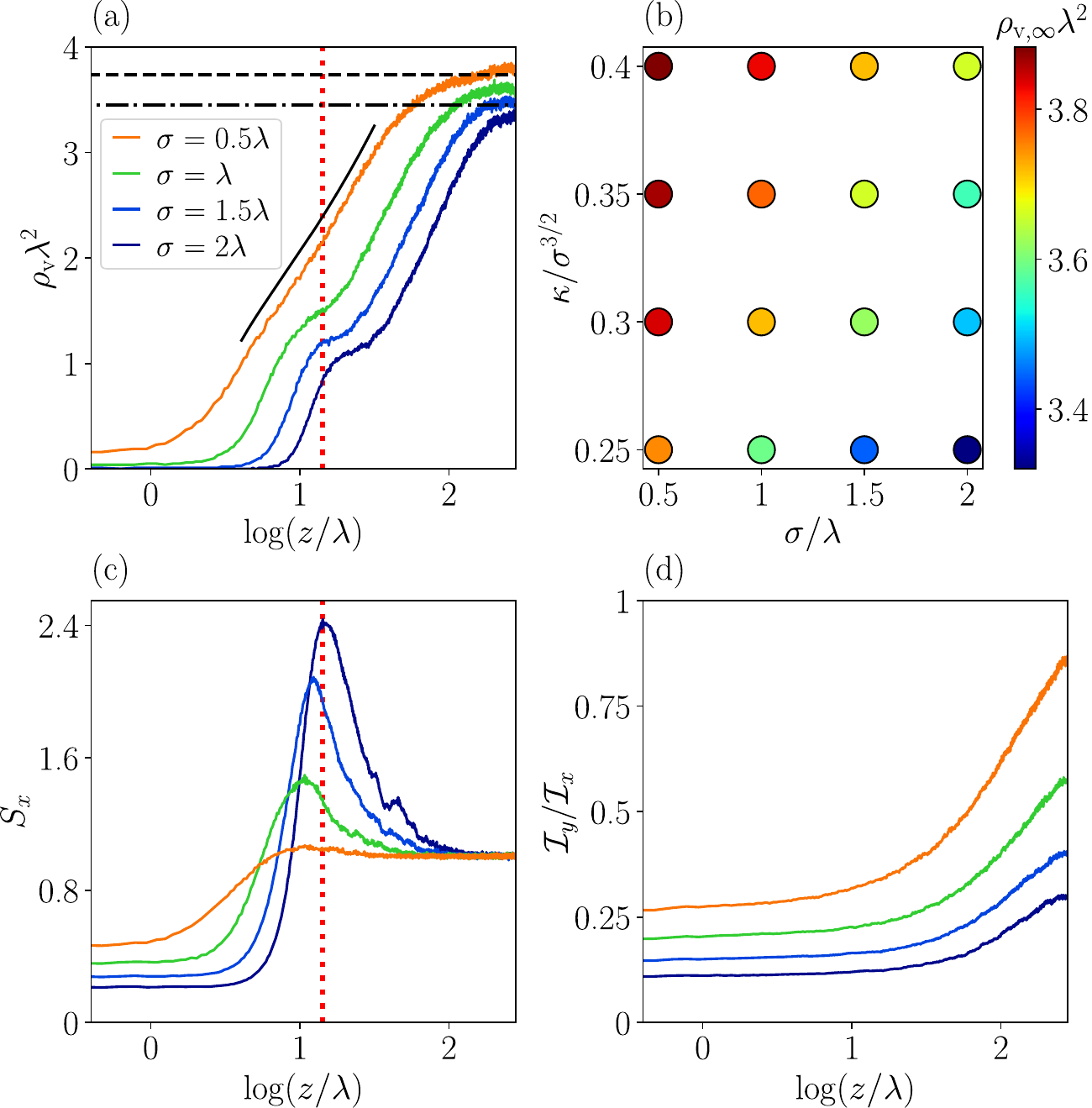}
\caption{\label{fig:vortex_density} (a) Evolution of vortex number per unit area $\rho_{\text{v}}$ for $\sigma = 0.5\lambda, \lambda, 1.5\lambda, 2\lambda$ and $\kappa = 0.25\sigma^{3/2}$, where the black solid line shows the trend for $0.5\lambda$. The dashed and dash-dotted horizontal lines are the estimations of $\rho_{\text{v}, \infty}$ obtained by applying two different filters to a random field in spatial Fourier domain. (b) Values of the asymptotic vortex density $\rho_{\text{v}, \infty}$ for various parameter sets of random media. In practice, use the vortex density at $z=280\lambda$ as asymptotic value. The points at the bottom row correspond to the four curves in panel (a) with the same coloring.
Panel (c) shows the evolution of the scintillation index for the four curves in panel (a), the maximum of which (e.g., vertical dotted line for $\sigma=2\lambda$) coinciding with the first branching point and the position where vortex nucleation is inhibited. 
Panel (d) shows the ratio of intensities in $y$ and $x$ polarizations for the four curves in panel (a). }
\end{figure}

We would like to proceed now with characterizing the dynamics using the number of vortices $N_{\text{v}}$. 
Here, we do not restrict ourselves to the weakly modulated  regime anymore. 
Figure~\ref{fig:vortex_density}(a) depicts the vortex density per unit area  $\rho_{\text{v}}=\lim_{\mathcal{A}\rightarrow\infty}N_{\text{v}}(\mathcal{A})/\mathcal{A}$ as a function of the propagation distance $z$. For that matter, we chose a set of realizations with fixed autocorrelation $C(0)$. 
The simulations use the same transverse domain size and discretization as in Fig.~\ref{fig:Hamiltonian}, with $L_z=320\lambda$ and $N_z=2048$.
We see that three stages of the dynamics become apparent. In the initial stage, where the paraxial approximation still holds, only a few vortices start to nucleate due to the accumulation of kinetic energy (cf.~Fig.~\ref{fig:Hamiltonian}). 

\begin{figure}[htbp]
\includegraphics[width=0.48\textwidth]{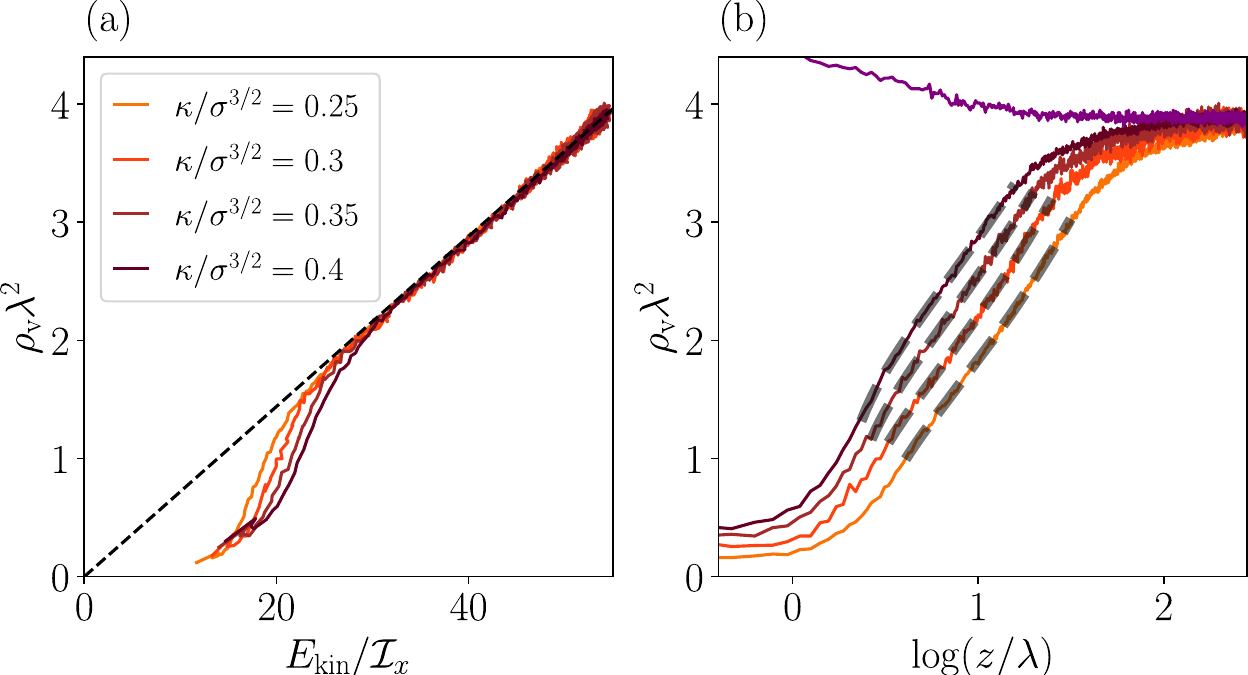}
\caption{\label{fig:vortex_kineticenergy} 
(a) The vortex density for $\sigma = 0.5\lambda$ as a function of the normalized kinetic energy $E_{\rm kin}/\mathcal{I}_x$ is approximately linear for sufficiently large vortex density. (b) The increase in vortex density as a function of the propagation distance can be approximated by $\rho_\text{v} \lambda ^2\propto (z/\lambda-z_{\rm cr}/\lambda)^\beta$. We find $\beta=0.36, 0.34, 0.30, 0.27$ for  $\kappa/\sigma^{3/2}=0.25, 0.3, 0.35, 0.4$. The purple line shows the propagation of the vortex density in the $x$ component for a source whose $x$ and $y$ components are both complex random fields (white noise) propagating in the random medium with $\kappa/\sigma^{3/2}=0.4$. 
}
\end{figure}

After that, there is an intermediate region, which features a more rapid increase in vortex number upon propagation. 
In case of sufficiently small correlation length ($\sigma=0.5\lambda$, orange line), we find that this increase can be approximated by a radical function with $\rho_{\text{v}}\lambda^2\propto (z/\lambda-z_{\rm cr}/\lambda)^\beta$, where we find $\beta\approx1/3$ [cf. Fig.~\ref{fig:vortex_kineticenergy}(b)]. 

The increase in vortex number appears to be inhibited for a short propagation interval if the correlation length is sufficiently large. The position where this thwarting occurs coincides with the point of maximum scintillation. The latter is shown in 
Fig.~\ref{fig:vortex_density}(c), where the scintillation index~\cite{Andrews:JOSAA:1999} is defined as $S_x(z) = \frac{\langle \mathcal{I}_x(z)^2\rangle}{\langle \mathcal{I}_x(z)\rangle^2}-1$.
This peak has been identified as a universal feature when studying propagation in random media in a classical context~\cite{Metzger:PRL:2010}, around which a delay in the creation of branches has been observed. This same universality appears in the number of vortices generated upon propagation as well. 
Upon further propagation, we find that although the random refractive index continues to drive the nucleation of more vortices, the dynamics leads to an equilibrium, a driven steady state where the nucleation and reconnection rates become equal. Thereby, we find a plateau of a roughly constant number of vortices upon propagation in Fig.~\ref{fig:vortex_density}(a). That is because the spatial spectrum of the wave can only broaden upon propagation until the evanescent limit is reached, and a further increase in complexity of the propagating wave function per unit area is not possible. 

We would like to contextualize the retrieved vortex density with what has been found before. The vortex density in scalar speckle fields in homogeneous media is commonly treated using Gaussian random-field models, i.e., superposition of many scalar plane waves with random phases~\cite{Goodman2015}.  Under this assumption, it has been found that the vortex density is inversely proportional to the coherence area~\cite{Freund:OptComm:1993, Freund:JoptSoc:94}. 
Furthermore, a prediction for the vortex density can be found assuming a radial power spectrum~\cite{Dennis:Proc:2000}. 

To estimate the asymptotic limit of the number of vortices per unit area for our case, we first generate a well-resolved complex scalar random field with Gaussian distributed random amplitude and uniformly distributed random phase in the range $[-\pi,\pi)$. Then, we apply a spectral filter $\sqrt{k_x^2+k_y^2}<\frac{2\pi n^*}{\lambda}$ to the random field in the Fourier domain to remove parts of the wave  that are beyond the evanescent limit. Since the refractive index of the random medium is not constant, it is not {\it{a priori}} clear what we ought to use for $n^*$. Employing 
$n^* = n_0$ leads to a cutoff radius that is spectrally too narrow, and thus we expect that using this cutoff will amount to a lower estimate for the predicted number of vortices per unit area. This is indeed what we observe in Fig.~\ref{fig:vortex_density}(a) (dash-dotted line). 
Alternatively, considering a spectrally broader interval by using $n^* =n_0+\sqrt{C(0)}$, leads to the (dashed line) and, unlike the first prediction, involves information about the correlation length and coupling strength. 
Both of these estimates neglect ``outliers" in spatial Fourier space, and thus
both seem to underestimate the number of vortices. 
However, by choosing random numbers that are equally likely in the whole spectral disk in Fourier space, we slightly overestimate the  complexity of the wave. 
These two estimates provide an interval within which the results of rigorous numerical computation can be expected to lie as can be seen in Fig.~\ref{fig:vortex_density}(a). 

Figure~\ref{fig:vortex_density}(b) depicts how the asymptotic number of vortices per unit area depends on the coupling strength $\kappa$ and correlation length $\sigma$ more systematically. We notice that an increase in coupling strength for a fixed correlation length, as well as a decrease in correlation length while keeping the autocorrelation $\kappa/\sigma^{3/2}$ fixed, consistently leads to an increase in the number of vortices. 
Both of these trends are not captured fully in the approximate model, as $n^* =n_0$ does not capture any dependency on $\kappa$ and $\sigma$. The approximation $n^* =n_0+\sqrt{C(0)}$ predicts that there should be no change for fixed autocorrelation $\kappa/\sigma^{3/2}$ (horizontal line). 
Figure~\ref{fig:vortex_density}(d) provides further context as to how vectorial the light becomes through scattering by displaying the ratio of the intensities $\mathcal{I}_y/\mathcal{I}_x$. 

We illustrate the connection between vortex density and kinetic energy normalized by the intensity $\mathcal{I}_x$ in Fig.~\ref{fig:vortex_kineticenergy}(a). The simulations use the same domain size and spatial discretization as in Fig.~\ref{fig:vortex_density}. We find that, whereas there is a region for which their relationship is not trivial, once the vortex density or kinetic energy is sufficiently large, there is a linear relationship between the two. This remains also true for other values of $\sigma$ (not shown). 
Figure~\ref{fig:vortex_kineticenergy}(b) shows that, indeed, the approximate cubic root relationship holds provided that $\sigma$ is sufficiently small. Alternatively, a reasonable alternative approximation is $\rho_\text{v}\lambda^2\propto \log(z/\lambda)$ (not shown). 

Let us now compare our findings with the 
case in which both $x$- and $y$-polarization components of the source ${\bm \xi}({\bm r})$ are initialized as complex random field (white noise) at the input $z=0$ plane. For this case, we consider the medium characterized by $\sigma=0.5\lambda$ and $\kappa/\sigma^{3/2}=0.4$. The resulting vortex density in $x$ component is shown by the purple line in Fig.~\ref{fig:vortex_kineticenergy}(b). We find that it converges to the same asymptotic value as a plane wave source polarized in the $x$ direction and propagating through the same medium. 
We observe that the field intensities asymptotically satisfy $\mathcal{I}_x=\mathcal{I}_y=\mathcal{I}_z$, indicating that the field becomes statistically isotropic with no preferred spatial direction. This explains why the two cases converge. 

\subsection{C. Kinetic energy spectrum}

Finally, we would like to characterize the kinetic energy further and study its internal distribution in terms of the inverse length scale or the spatial Fourier wave vector. Similar to quantum turbulence theory~\cite{Bradley:PRX:2012,Glorieux:PRA:2023}, a more detailed characterization can be achieved by studying the energy spectrum. 
The total kinetic energy of a field $E_x = A\mathrm{e}^{\mathrm{i}\Phi}$ can be decomposed as 
$
    \int |\nabla_\perp E_x|^2 d^2r_\perp = \int |\nabla_\perp A|^2 + |A\nabla_\perp\Phi|^2 d^2r_\perp.
$ Here, $\nabla_\perp \Phi$ represents the ``velocity" of the flux in the transverse plane. Its value increases near the phase singularity. 
We introduce now the amplitude-weighted velocity field as ${\bm  u}=A\nabla_\perp \Phi$. For context, similar definitions have been employed; e.g., in Refs.~\cite{Dennis:Proc:2000, Zhang:PRL:2007, Zhang_Genack:PRL:2007}, the intensity-weighted velocity field has been considered. 
The amplitude-weighted velocity field can be further decomposed into a compressible and an incompressible part using Helmholtz decomposition (e.g., Ref.~\cite{Maucher:PRL:2018}), $ {\bm  u} = {\bm  u}^{\text{c}} + {\bm  u}^{\text{i}}$.  
This can be expressed in components as $u_i=\partial_i f+\epsilon_{ij}\partial_j g=u^\text{c}_i+u^\text{i}_i$, where $f$ and $g$ are scalar functions. 
By construction, there is an incompressible ${\bm  u}^{\text{i}}$ or divergence-free part of the field ${\bm  u}$ that satisfies $\partial_i u_i^{\text{i}} = 0$, as well as a compressible or curl-free part ${\bm  u}^{\text{c}}$ satisfying $\epsilon_{ij}\partial_i u_j^{\text{c}}=0$. 

The kinetic energy of the incompressible part is defined as 
$    E_\text{kin}^\text{i} = \int |{\bm  u}^\text{i}({\bm  r}_\perp)|^2d^2{\bm  r}_\perp$.
It is useful to find its dependence on the modulus of the wave number $k$ by integrating the incompressible field in Fourier domain over the azimuthal angle in polar coordinates: 
\begin{equation}
    E_\text{kin}^\text{i}(k) = k\int_0^{2\pi}|\hat{\bm  u}^\text{i}({\bm k})|^2 d\phi_k, 
\end{equation}
where $\hat{\bm  u}^\text{i}({\bm k}) = \mathcal{F}_\text{2d}\{{\bm  u}^\text{i}({\bm  r}_\perp)\}$ is the two-dimensional Fourier transform of the incompressible field. 
In case of sufficiently large $\sigma$ and sufficiently small $\kappa$ when the wave propagation becomes effectively paraxial, the tail of the incompressible kinetic energy can be expected to mimic that of a single vortex. 
To match the two, we consider a finite single vortex with a Gaussian profile and the amplitude being a free parameter. 
Hence, we assume a quasilinearly polarized beam with profile $E_x=r_\perp\mathrm{e}^{-\frac{r_\perp^2}{\sigma^2}}\mathrm{e}^{\pm\mathrm{i}\theta}$ with $\theta = \arctan(y/x)$. In that case, it is simple to find the incompressible kinetic energy as  
\begin{equation}
    E_\text{kin}^\text{i}(k) \propto \sigma^6k^3\mathrm{e}^{-\frac{\sigma^2k^2}{4}}\left[I_0 \left(\frac{\sigma^2k^2}{8}\right)-I_1\left(\frac{\sigma^2k^2}{8}\right)\right]^2\nonumber, 
\end{equation}
where $I_j$ is the modified Bessel function of the first kind of order $j$. Thus, we chose the correlation length $\sigma$ as the width of the Gaussian envelope. 
This expression asymptotically scales as $\sim k^{-3}$, and the peak is inversely proportional to length scale $\sigma$. 

\begin{figure}[htbp]
\includegraphics[width=0.48\textwidth]{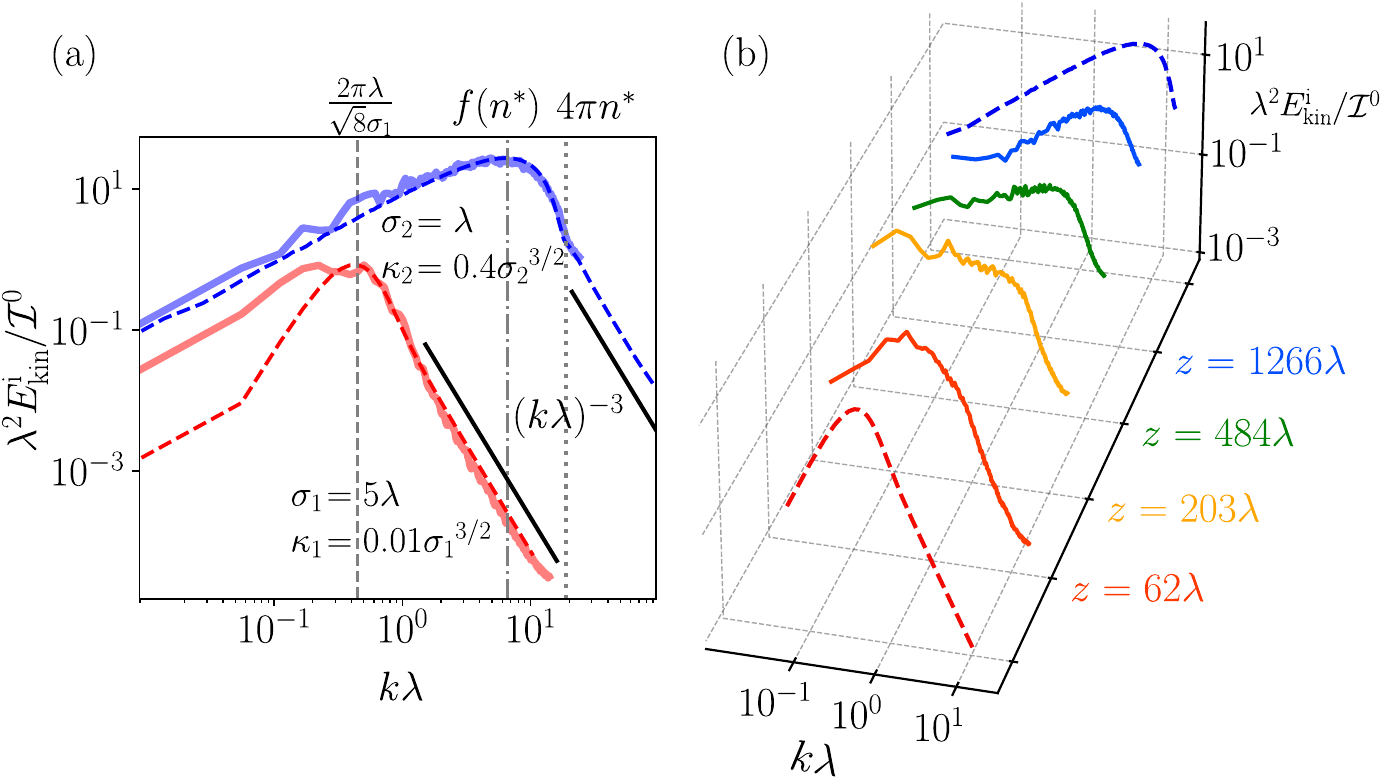}
\caption{\label{fig:energy_cascade} (a) The incompressible kinetic energy as a function of the wave vector $k$ for a paraxial case (red) and a vectorial, nonparaxial case (blue). The dashed lines correspond to the theoretical models and the solid lines to the numerics. The black solid lines show the $k^{-3}$ ultraviolet asymptotics. (b) The dashed lines correspond to the theoretical calculations in panel (a), and the solid lines show the spectra upon propagation, showcasing how an initial plane wave moves from one limiting paraxial case to the random field. }
\end{figure}

The comparison between theory and rigorous numerics of the incompressible kinetic energy spectrum for the paraxial case is shown in Fig.~\ref{fig:energy_cascade}(a) for the case $\sigma_1 = 5\lambda$ and $\kappa_1 = 0.01\sigma_1^{3/2}$. The simulation uses the same domain size and spatial discretization as in Fig.~\ref{fig:vortex_density}. To obtain this spectrum, we propagated an initially linearly polarized plane wave to $z=310\lambda$.  
To compare numerics to theory, we multiplied the theoretical incompressible kinetic energy with a constant to match the peak value of the numerical one. 
We find that, indeed, the tail as well as the peak position $k_{\text{peak}}$ is well predicted by the analytical model of the single vortex. The peak corresponds to the inverse length scale at which energy is injected, i.e., proportional to $1/\sigma_1$. In fact, the analytical prediction yields $k_{\text{peak}}=2\pi/(\sqrt{8}\sigma_1)$, shown as the dashed gray vertical line in Fig.~\ref{fig:energy_cascade}(a). 

Let us now consider the situation where the wave is nonparaxial and vectorial, shown in Fig.~\ref{fig:energy_cascade}(a) at a propagation distance where the driven steady state has been reached with $\sigma_2=\lambda$ and $\kappa_2=0.4\sigma_2^{3/2}$. It is numerically challenging to show the ultraviolet range up to the point where the asymptotic behavior $k^{-3}$ becomes obvious. 
For achieving that, we used 128 cores and a grid of $1024^3$ points ($L_x=L_y=L_z=160\lambda$) and benchmarked against $512^2\times2048$ ($L_x=L_y=80\lambda, L_z=320\lambda$). 
We would like to understand the numerical result (solid blue line) employing a suitable model (dashed blue line). To proceed, we use the method employed before to obtain a prediction for the number of vortices per unit area again by considering a random field and spectrally filtering out the evanescent part. 
Note that in this case the peak becomes completely independent of $\sigma_2$, and the only relevant length scale that matters for the theoretical model is $\lambda/n^*$. Here, we used $n^*=n_0$. The theoretical model (blue dashed line) features a kink at large values of $k$, which is located at $k = 4\pi n^*/\lambda$ and is due to the spectral filtering. The theoretical result is simulated using a two-dimensional grid of $2^{14}\times2^{14}$ points with a domain size $320\lambda\times320\lambda$ and averaged over 100 realizations. 
We find that there is good agreement with the actual numerical data acquired through propagation, including the pronounced kink due to evanescent waves. After the kink, the theoretical spectrum features the expected ultraviolet $k^{-3}$ behavior. 

We now reinspect the case of $\sigma=5\lambda$, increase $\kappa$ to $\kappa=0.1\sigma^{3/2}$, and aim at obtaining a  qualitative understanding of the dynamics of the evolution of the spectra. This is shown in Fig.~\ref{fig:energy_cascade}(b). 
The simulation uses the same transverse domain size and discretization as in Fig.~\ref{fig:Hamiltonian} ($L_x=L_y=80\lambda, N_x=N_y=512$), with $L_z=1280\lambda$ and $N_z=8192$.
The initial and final dashed curves correspond to the models (dashed curves) shown in Fig.~\ref{fig:energy_cascade}(a). 
At an early stage ($z=62\lambda$), where only a few vortices nucleated, the numerical spectrum (red solid line) resembles the spectrum of the paraxial prediction (red dashed line). Upon further propagation, the peak is blueshifted, i.e., to larger values of $k$. Ultimately, at $z=1266\lambda$, the spectrum features the characteristic distribution of the evanescently filtered random spectrum - shown as the blue dashed line in panel (b), which is equivalent to panel (a) as we chose $n^*=n_0$. 

\section{IV. Conclusions}

In summary, we presented a statistical study of the kinetic energy of the transverse flux of light propagating in random media at the paraxial-nonparaxial interface. 
The expressions for the paraxial case are valid for weakly modulated refractive index media, such as beam propagation in turbulent air flow in atmospheric physics~\cite{Fante:ProcIEEE:1975} or in Schlieren imaging~\cite{Settles_2001}. 

Upon exploring the nucleation dynamics of vortices, we identified the following features: First, after an initial period where barely any vortices nucleate, there is a point at which, for sufficiently small correlation length, we find a scaling of vortex density of roughly $\rho_{\text{v}}\lambda^2\sim (z/\lambda)^{1/3}$.
Second, there is a kink in the nucleation rate if the correlation length is sufficiently large, similar to what has already been classically observed for the amplitude distribution~\cite{Metzger:PRL:2010}. 
Third, we found that ultimately the nucleation and reconnection rates of vortices become equal and that no further complexity can be added to the field due to evanescence and found a semianalytical estimate for the latter. 

Furthermore, we examined the distribution of optical incompressible kinetic energy as a function of the wave vector with attention to the paraxial - nonparaxial transition. 
We found that continuously injecting energy at the inverse correlation length first leads to a spectrum that has the peak at that same inverse length. However, upon further propagation this peak blueshifts and acquires the shape of an evanescently filtered random spectrum. Thereby, we gained an understanding of the emergence of the driven steady state presented in Fig.~\ref{fig:vortex_density} and the internal distribution of length scales that contribute to the kinetic energy. Furthermore, we provided a means for the characterization of the transition to a nonparaxial, vectorial distribution of light. 
The dynamics is different from what is usually discussed in fluid dynamics and quantum fluids, where energy cascades emerge upon injecting energy and dissipate at specific scales until a steady state with its characteristic slope is developed due to nonlinearity. 

In this paper, only one length scale is fixed, i.e., the correlation length to reduce the parameter space. Therefore, this represents an effective model. Generalizing our findings to include multiple length scales, present when considering, e.g., scatterers with given mean distance and radii, appears to be a further natural extension. 
Adding nonlinearity for establishing a cascade has already been undertaken in Ref.~\cite{Glorieux:PRA:2023} in the paraxial regime. Extending that to the nonparaxial regime and establishing a cascade in that case appears to be a further interesting avenue.
This could be combined with nonlinear processes, such as second harmonic generation in nonlinear random media~\cite{Grange:PRR:2025}. 

\section{Acknowledgment}
We would like to thank the  Department of Precision and Microsystems Engineering for financial support.

%\bibliography{apssamp}
%apsrev4-2.bst 2019-01-14 (MD) hand-edited version of apsrev4-1.bst
%Control: key (0)
%Control: author (8) initials jnrlst
%Control: editor formatted (1) identically to author
%Control: production of article title (0) allowed
%Control: page (0) single
%Control: year (1) truncated
%Control: production of eprint (0) enabled
%

\end{document}